\documentclass{emulateapj}
\usepackage{xspace}
\usepackage{amsmath}
\usepackage{framed}
\usepackage{txfonts}
\usepackage{epstopdf} 
\def\figdir{.}

\def\H2{{{\rm H}_2}}
\def\HI{{\rm H\,I}}

\def\Msun{\, {\rm M}_{\odot}}

\def\dim#1{\mbox{\,#1}}
\def\figname#1#2{\figdir/#1}

\def\hide#1{}

\begin{document}

\title{Effect of Cosmic UV Background on Star Formation in High Redshift Galaxies}

\author{Nickolay Y.\ Gnedin\altaffilmark{1,2,3}}
\altaffiltext{1}{Particle Astrophysics Center, 
Fermi National Accelerator Laboratory, Batavia, IL 60510, USA; gnedin@fnal.gov}
\altaffiltext{2}{Department of Astronomy \& Astrophysics, The
  University of Chicago, Chicago, IL 60637 USA} 
\altaffiltext{3}{Kavli Institute for Cosmological Physics, The University of Chicago, Chicago, IL 60637 USA} 

\begin{abstract}
The effect of the cosmic UV background on star formation in high
redshift galaxies is explored by means of high resolutions
cosmological simulations. The simulations include star formation, 3D
radiative transfer, and a highly detailed ISM model, and reach spatial
resolution sufficient to resolve formation sites for molecular
clouds. In the simulations the local radiation field in the
Lyman-Werner band around star-forming molecular clouds dominates over
the cosmic UV background by a factor of 100, similarly to the
interstellar radiation field in the Milky Way and in a few high
redshift galaxies for which measurements exist. The cosmic UV
background, therefore, is essentially irrelevant for star formation in
normal galaxies.
\end{abstract}

\keywords{cosmology: theory -- galaxies: evolution -- galaxies:
  formation -- stars:formation -- methods: numerical}

\section{Introduction}
\label{sec:intro}

Since the end of cosmic reionization, the intergalactic space has been
filled with the accumulated ultra-violet radiation from the previous
generations of massive stars and quasars, the so-called cosmic UV
background. The UV background completely controls the ionization
state of the intergalactic medium (IGM) that manifests itself in the
numerous Ly-$\alpha$ absorption lines in the spectra of distant
quasars, the Ly-$\alpha$ forest \citep[see][for a recent review]{igm:m09}.

In the interstellar medium of local galaxies (ISM) the situation
appears to be reversed: the Haardt-Madau model for the UV background
\citep{jnu:hm01} predicts the $z=0$ radiation field at $1000\AA$ of
about
$2\times10^3\dim{photons}/\dim{cm}^2/\dim{s}/\dim{ster}/\dim{eV}$,
some 500 smaller than the interstellar radiation field in the Milky
Way at the solar circle \citep{h2:d78,h2:mmp83}.

At higher redshift the situation is, however, less clear. For example,
the UV background at $z\sim3$ is expected to be 50 to 100 times higher
than at $z=0$ \citep{jnu:hm01,jnu:flzh09}. On the other hand,
estimates of the interstellar radiation field in high redshift
gamma-ray burst hosts also give some 100 times higher values
than the Milky Way field \citep{hizgal:cppp09}, but the observational
constraints remain sparse.

In the cosmological simulation community, the effect of the cosmic UV
background is often included in the cooling rates
\citep[e.g.][etc]{sims:co92,sims:kwh96,sims:ns97,sims:k03,sims:ck09,sims:svbw09,sims:honj09}. However,
the effect of the local interstellar radiation field (that dominates
over the cosmic UV background by a larger factor at least at $z=0$,
and may be at high redshift too) has not yet been included in those
simulations. In this paper the effect of the cosmic UV background is
critically reassessed with numerical simulations that both include the
full 3D, time-dependent and spatially variable treatment of radiative
transfer and have high enough spatial resolution to resolve the sites
of molecular clouds and associated star formation.

\section{Simulations and star formation model}
\label{sec:sims}

The physical ingredients and computational setup for the simulations
used in this paper have been recently described in great detail
elsewhere \citep{ng:gk10a,ng:gk10b}. As a brief reminder, the
simulations have been performed with the Adaptive Refinement Tree (ART) code
\citep{misc:k99,misc:kkh02,sims:rzk08} that uses adaptive mesh
refinement in both the gas dynamics and gravity calculations to
achieve high dynamic range in spatial scale.

The simulations include star formation and supernova enrichment and
thermal energy feedback, as well as a highly detailed ISM model. The
3D radiative transfer of UV radiation from individual stellar
particles is followed self-consistently with the OTVET approximation
\citep{ng:ga01}. The simulations incorporate non-equilibrium chemical
network of hydrogen and helium and non-equilibrium,
metallicity-dependent cooling and heating rates, and a
phenomenological model of molecular hydrogen formation on and
shielding by cosmic dust, as well as $\H2$ self-shielding
\citep{ng:gk10a,ng:gk10b}.

Particular simulations used in this paper model a small region
including a Milky-Way progenitor galaxy and a number of smaller
galaxies with the mass resolution of $1.3\times10^6\Msun$ in dark
matter, $2.2\times 10^5\Msun$ in baryons, and with the spatial resolution of
$65\dim{pc}\times[4/(1+z)]$ (in physical units) within the fully
refined region.

Star formation in the simulation is occurring in the molecular gas
only, using the prescriptions of \citet{sfr:km05} and
\citet{sfr:kt07}. The exact formulation of the star formation recipe
is shown in Equation (2) of \citet{ng:gk10b}.  

For the purpose of this paper, two simulations are considered that
differ only by the inclusion of the cosmic UV background from
\citet{jnu:hm01}. In the simulation without the cosmic UV background,
the local radiation field produced by nearby massive stars is still
included in exactly the same manner as in the simulation with the UV
background. Thus, these two simulations can be used to evaluate the
particular effect of the cosmic UV background on the properties of 
model galaxies. Because of computational expense, the simulations are
not continued beyond $z=2$.

The simulation with the cosmic UV background is the same one as
described in \citet{ng:gk10a}. That reference also shows good agreement of
that simulation with several observational constraints on the
properties of high redshift galaxies.

\section{Results}
\label{sec:res}

\begin{figure}[t]
\includegraphics[scale=0.43]{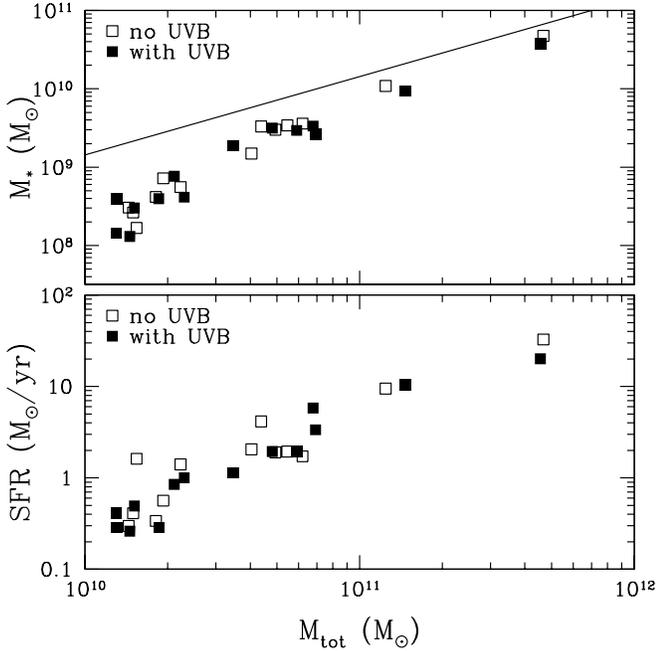}
\caption{The stellar mass (top) and the star formation rate (bottom)
  vs the total mass for all fully resolved model galaxies at
  $z=3$. Open squares show the run without the cosmic UV background,
  while filled squares show the fully self-consistent run, with the UV
  background.  The star formation rates are averaged over
  $20\dim{Myr}$.\label{fig:gals}}
\end{figure}

A direct comparison of stellar masses and star formation rates between
the galaxies in the two simulations is shown in Figure
\ref{fig:gals}. In order to minimize the effect of finite numerical
resolution, only highly resolved galaxies - i.e.\ galaxies that reach
the 8th level of mesh refinement ($130\dim{pc}$ spatial resolution at $z=3$)
within their gaseous disks - are shown in this and all subsequent
figures. At this spatial resolution the sub-cell model for $\H2$
formation performs reliably, as is demonstrated in Figure 13 of
\citet{ng:gk10b}. Galaxies with masses below $M_{\rm tot}=10^{10}\Msun$ are not
sufficiently resolved in these simulations - none of such galaxies
achieves 8 levels of mesh refinement within their gaseous disks.

While the two simulations do not produce identical
results, the difference between the two runs is not dramatic and is
fully consistent with the timing differences in two simulations that
have slightly different time-steps and, hence, times of intermediate
outputs. Thus, no significant effect of the cosmic UV background on
the global properties of simulated galaxies is observed in these
simulations. 

\begin{figure}[t]
\includegraphics[scale=0.43]{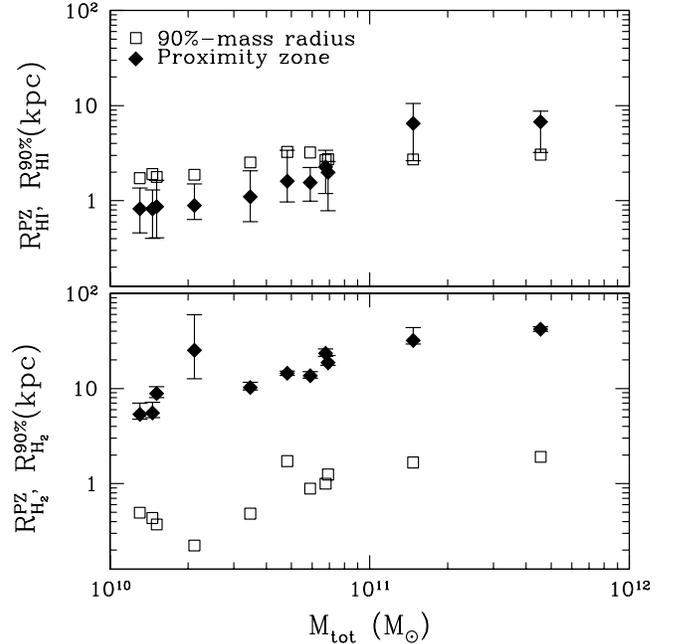}
\caption{Comparison between the proximity zones (filled diamonds) and
  radii containing 90\% of the neutral (top) and molecular (bottom)
  gas in model galaxies at $z=3$ (open squares). Vertical error-bars
  give 10\% - 90\% range for the proximity zone sizes as viewed in
  different directions from the center of each galaxy.
  \label{fig:pzcomp}}
\end{figure}

The origin for this conclusion becomes apparent from Figure
\ref{fig:pzcomp}, where the sizes of the proximity zones -
regions around model galaxies that are dominated by the local
radiation from the galaxies themselves and not by the cosmic UV
background - are shown together with radii containing most of atomic
and molecular gas.

A size of a proximity zone depends, of course, on the wavelength of
radiation, because, in general, spectral shapes of the local
interstellar radiation field and the cosmic UV background are
different. Figure \ref{fig:pzcomp}, therefore, shows two important
wavelengths: the Lyman limit and $\lambda=1000\AA$, in the middle of
the Lyman-Werner band. Radiation above the Lyman limit ionizes neutral
hydrogen, and so is important for determining which part of the
galactic gas can cool rapidly. Radiation in the Lyman-Werner band
destroys molecular hydrogen; only when gas is shielded from that
radiation by cosmic dust and molecular hydrogen self-shielding can it
become fully molecular (and, hence form stars).

In a given frequency band, absorption of radiation emitted by stars in
a galaxy will, generally, be different in different
directions. Therefore, the proximity zone of a given galaxy is not
necessarily spherical, but can have a complex shape.  In order to
quantify variation in the shapes of proximity zones, a HEALPix
tessellation \citep{misc:ghbw05} of the celestial sphere (as seen from
the center of a particular galaxy) is constructed and the proximity
zone size is measured separately for each pixel on the sky. Because
HEALPix provides uniform sampling of all possible directions, it
offers a convenient way to quantify variations in sizes of proximity
zones in different wavebands. Error-bars in Figure \ref{fig:pzcomp}
show the 10\% - 90\% range in the size of the proximity zone as a
function of the angle on the sky.

To evaluate the effect of the local radiation field on the gas
physics, Figure \ref{fig:pzcomp} also displays radii that contain 90\%
of atomic (top) and molecular (bottom) gas in the model galaxies. As
can be seen, the proximity zone in the Lyman-Werner band extends over
10 times beyond the edge of the molecular gas. That explains why the
star formation in the simulations is insensitive to the presence of
the cosmic UV background - the local radiation field in the sites of
star formation (i.e.\ in molecular clouds) always dominates over the
cosmic background by a factor of about 100.


The situation appears to be reversed for atomic hydrogen. The escape
fraction for ionizing radiation is small in these simulations,
consistent with observational estimates \citep{ng:gkc08}. It is
particularly small along the disk, so that little local ionizing
radiation shines on the outer parts of $\HI$ disks; instead, the edges
of $\HI$ disks are determined by the cosmic background. 

\begin{figure}[t]
\includegraphics[scale=0.43]{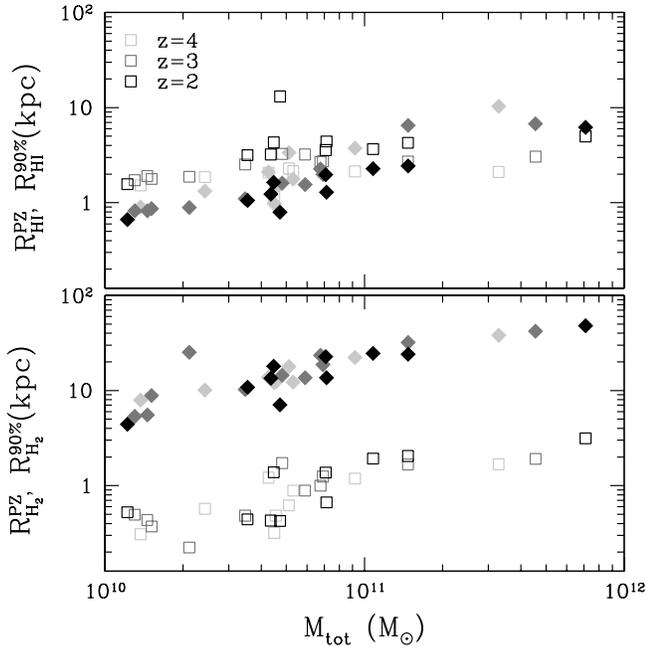}
\caption{Comparison between the proximity zones and radii containing
  90\% of the neutral and molecular gas for the fully self-consistent
  run, with the UV background, at three different redshifts. Symbol
  and panels are similar to Figure \ref{fig:pzcomp}. Three shades of
  gray show three different redshifts, as marked in the figure.
\label{fig:pzevol}}
\end{figure}

Finally, to verify the robustness of these results,  Figure
\ref{fig:pzevol} presents proximity zones and 90\%-mass radii for the full
self-consistent simulation (with the cosmic UV background) at 3 different
redshifts. While there is substantial scatter between individual
points, the trends observed at $z=3$ are reproduced at other redshifts
as well.

\section{Conclusions}
\label{sec:discussion}

The results of this paper can be summarized in just a few lines: the
local radiation field in the Lyman-Werner band around star-forming
molecular clouds dominates over the cosmic UV background by a factor
of 100, similarly to the interstellar radiation field in the Milky Way
\citep{h2:d78,h2:mmp83} and in a few high redshift galaxies for which
measurements exist \citep{hizgal:cppp09}. The cosmic UV background,
therefore, is essentially irrelevant for star formation in normal
galaxies. Only in the lowest mass dwarfs, where the radiation can
actually photo-evaporate the ISM, may the cosmic background affect
star formation. Such galaxies, however, contribute very little to the
total star formation history of the universe. 

The situation is more complex for ionizing radiation - in that band
the cosmic UV background may play an important role in determining the
exact locations of the edges of extended $\HI$ disks, in agreement
with the earlier results of \citet{igm:s04a}. Those outside regions of
galactic disks, however, contain little molecular gas and, therefore,
are inert to star formation.

The simulations presented here do not continue beyond $z=2$, and so
can not be used for testing the results of \citet{sims:honj09}, who
found little effect of the cosmic UV background on star formation in
normal galaxies at $z>2$, but a larger effect at lower
redshifts. Simulations of \citet{sims:honj09}, however, did not
include radiative transfer and, therefore, failed to account for the
local radiation and its dominance over the cosmic UV background within
the galactic proximity zones. For example, if the Milky Way galaxy is
representative of low redshift ($z<2$) normal galaxies, then one may
expect that even at low redshifts local radiation dominates over the
cosmic background within the galactic ISM, rendering the background
irrelevant to star formation.

\acknowledgements 

I am grateful to Andrey Kravtsov for comprehensive comments and
enlightening discussion. This work was supported in part by the DOE at
Fermilab, by the NSF grant AST-0908063, and by the NASA grant
NNX-09AJ54G. The simulations used in this work have been performed on
the Joint Fermilab - KICP Supercomputing Cluster, supported by grants
from Fermilab, Kavli Institute for Cosmological Physics, and the
University of Chicago.  This work made extensive use of the HEALPix
spherical tessellation package
and the NASA Astrophysics Data System and
{\tt arXiv.org} preprint server.

\bibliographystyle{apj}
\bibliography{ng-bibs/misc,ng-bibs/sfr,ng-bibs/self,ng-bibs/sims,ng-bibs/igm,ng-bibs/jnu,ng-bibs/h2,ng-bibs/hizgal}

\end{document}